\documentclass[showpacs,preprintnumbers,a4paper,aps,12pt]{revtex4}
\usepackage{bm}
\usepackage{graphicx}
\usepackage{amssymb}
\usepackage{amsmath}
\usepackage{hyperref}

\usepackage{epsfig}
\newcommand{\be}{\begin{equation}}
\newcommand{\ee}{\end{equation}}
\newcommand{\bd}{\begin{displaymath}}
\newcommand{\ed}{\end{displaymath}}
\newcommand{\bea}{\begin{eqnarray}}
\newcommand{\eea}{\end{eqnarray}}

\begin{document}

\title{ First Principles Investigation of Hydrogen Physical Adsorption on
  Graphynes' layers }


\author{Massimiliano Bartolomei}\email{maxbart@iff.csic.es}
\affiliation{Instituto de F\'{\i}sica Fundamental,
Consejo Superior de Investigaciones Cient\'{\i}ficas (IFF-CSIC), Serrano 123,
28006 Madrid, Spain}

\author{Estela Carmona-Novillo}
\affiliation{Instituto de F\'{\i}sica Fundamental,
Consejo Superior de Investigaciones Cient\'{\i}ficas (IFF-CSIC), Serrano 123,
28006 Madrid, Spain}

\author{Giacomo Giorgi}
\affiliation{Department of Chemical System Engineering, School of Engineering,
University of Tokyo, Tokio, Japan}



\begin{abstract}
Graphynes are 2D porous structures deriving from graphene featuring triangular and regularly
distributed subnanometer pores, which may be exploited to host small gaseous species. 
First principles adsorption energies of molecular hydrogen (H$_2$) on graphene,
graphdiyne and graphtriyne molecular prototypes are obtained at the MP2C level of theory.
First, a single layer is investigated  and it is found that graphynes are more
suited than graphene for H$_2$ physical adsorption since they provide
larger binding energies at equilibrium distances much closer to the 2D
plane. In particular, for graphtriyne a flat minimum located right in the
geometric center of the pore is identified.
A novel graphite composed of graphtriyne stacked sheets is then proposed and
an estimation of its 3D arrangement is obtained at the DFT
level of theory. In contrast to pristine graphite this new carbon material
allow both H$_2$ intercalation and out-of-plane diffusion by exploiting the
larger volume provided by its nanopores. Related H$_2$ binding energies for
intercalation and in-pore adsorption are around 0.1 eV and they could lead to
high storage capacities.
The proposed carbon-based layered material could represent a safer and potentially cheaper 
alternative for hydrogen on-board storage than conventional solutions based on
cryogenic liquefaction and/or high compression.
\end{abstract}

\date{\today}

\maketitle


\section{Introduction}


Hydrogen is a clean-burning energy carrier which is supposed to provide 
the most promising alternative to fossil fuels to be employed in
propulsion systems for automotive applications. 
However, some drawbacks such as the safe and convenient
hydrogen storage\cite{reviewH2storage:04} and release in reversible cycles still
represent major challenges to the scientific community.  
The use of advanced solid state adsorbents has recently received wide
attention since they could provide solutions  to achieve the 2017 targets
  for on-board vehicle hydrogen storage established by the
  U.S. Department of Energy (DOE): they are actually  5.5 wt\%  and 0.040
  kg\,L$^{-1}$ for gravimetric and volumetric
capacities, respectively\cite{DOE-note}, and they have not been simultaneously reached
yet in practice at moderate temperature and pressure.

As a matter of fact, the physisorption of molecular hydrogen (H$_2$) in porous
materials represents a  safer, simpler and potentially cheaper option for
gas storage than conventional solutions, based on liquefaction at low
temperature (and/or high compression at room temperature). Materials
traditionally used in these applications include activated
carbons and zeolites and in recent years novel and promising adsorbents, namely metal-organic
frameworks (MOF) and covalent organic frameworks (COF)\cite{MRS:13}, are also available. 
MOF and COF possess advantages with respect to traditional porous
materials due to their crystalline structure: in fact, their regular and
ordered porosity, together with tunable size and shape of the openings, has
led to very high performances in gas uptake
capacities\cite{ChemSocRev:12}. However, they
 present some shortcomings mainly related to their thermal and
chemical stability. 
Moreover, the presence of heavy metals in MOFs, which in general enhances the
binding strength of a given adsorbate, could also induce its dissociation
which would alter the desirable gas storage and delivery reversible cycle.
Therefore, investigations on alternative porous materials are highly advisable
and, specifically, we here focus on carbon-based layers, which in general provide a large 
 specific surface area coupled with a lower weight. 

In the last years, by following ``bottom-up'' assembly processes
two-dimensional  (2D) materials similar to ubiquitous graphene but with
regular and uniformly distributed subnanometer pores have
been synthesized in large area films\cite{Bieri:09,graphd-chemcomm:2010},  and
among them  graphdiyne is of particularly interest
for our purposes.
 Graphdiyne is actually a member of $\gamma$-graphynes which are new 2D carbon allotropes formed by sp-sp$^2$
hybridized carbon atoms. They can be thought as deriving from graphene
 by replacing one-third of its C-C bonds with mono(poly)-acetylenic units.
The number {\it n} of acetylenic linkages which connect the benzene
rings defines the different graphyne-{\it n} species and the first three members of
the family are known as graphyne, graphdiyne and graphtriyne,
respectively\cite{narita:98}, and they feature nanopores of increasing size.  
 The successful synthesis of graphynes has led to important theoretical
studies devoted to their application as effective single-layer
membranes for gas separation and water filtration
technologies\cite{nanoscalemitbis:2012,nanoscalemit:2013,jpclours:2014,jpccours:2014}.

In this work we want to assess the capability of graphynes
 as efficient materials for the reversible storage of molecular hydrogen
and a  fundamental point is to address the possibility of gas
adsorption by exploiting its intercalation.
 As it is well known intercalation of gases between pristine graphene layers
 it is not feasible since large values of the interlayer distance would be
 needed for an effective storing\cite{PNAS:05,Alonso:2008} (roughly the double with respect to graphite
 equilibrium distance). This requirement is difficult to fulfill in practice
 since it depends on an ``a priori'' forced separation of the graphene layers.
 On the contrary novel graphites composed of  multi-layer graphynes could allow
 intercalation without significantly altering the interlayer separation since a larger
 available volume is in principle provided by the naturally occurring
 nano-pores.

The interaction of nonpolar H$_2$ molecules with carbon-based
substrates is mainly the London dispersion and expected values for its binding
energy with a single(multi)-layer are around 50 meV\cite{Vidali:91,dft-cc:10,dft-vdw:14}.
The H$_2$--graphdiyne single-layer interaction was previously theoretically
investigated\cite{graphd-chemcomm:2011,nanoscalemitbis:2012} by means of
dispersion-corrected density functional theory (DFT)
approaches but the focus was mostly on the penetration barrier
 in order to propose graphdiyne  as an optimal platform for hydrogen
purification.
The three-dimensional (3D) diffusion of H$_2$ in bulk graphdiyne was also recently
investigated\cite{h2-graphd:2013} but the obtained dispersion-corrected DFT interaction energies  
 were unreasonably too high with binding energies larger than 400 meV.
More  accurate estimations for the  noncovalent H$_2$--graphynes 
 interaction are therefore desirable and their calculations would
 require reliable computational approaches. Our choice is to use the 
 ``coupled'' supermolecular second-order M{\o}ller-Plesset perturbation theory
(MP2C) which has been reported to provide reliable estimations for weakly bound systems
such as rare gas--fullerene\cite{mp2c-full} and  -coronene\cite{grapheneours:2013}
as well as water(rare gas)--graphynes' pores\cite{jpclours:2014,jpccours:2014}.

The work reported in this paper is based on the following scheme.
First, we carry out accurate estimations of the interaction energy of H$_2$
adsorbed on graphene, graphdiyne and graphtriyne molecular prototypes. 
Then, after assessing the most appropriate
graphyne for H$_2$ physisorption, we obtain the equilibrium 3D structure of
the related graphyne multi-layer.
Finally, the H$_2$ interaction with the proposed novel graphite composed of 
graphyne layers is also investigated.


\section{Computational Methods}
\label{sec.2}

The electronic structure calculations for the H$_2$ adsorption energy
 have been carried out at the
MP2C\cite{mp2c} level of theory by using the Molpro2012.1 package\cite{MOLPRO}.
For the graphene prototype the C-C bond length is 1.42 \AA\,
  while for the graphyne pores we have considered the following bond lengths\cite{mech-graphdiyne:2012}: 
1.431 \AA\, for the aromatic C-C, 1.231 \AA\,
for triple C-C, 1.337 \AA\, for the single C-C between two triple C-C bonds,  
1.395 \AA\, for the single C-C connecting aromatic and triple C-C bonds. In
all cases the C-H and H-H bond lengths are 1.09 and 0.746 \AA, respectively.
 The aug-cc-pVTZ\cite{Dunning} basis set was employed for the molecular planar structures, 
while the aug-cc-pVQZ\cite{Dunning} basis has been used for the H$_2$
molecule. 
All considered molecular structures are treated as rigid bodies: the atoms composing the
investigated graphene and graphynes prototypes are frozen in their initial positions and the
molecular configuration of H$_2$ is not allowed to relax during the calculations.  
The interaction energies have been further corrected for the basis set
superposition error by the counterpoise method of Boys and
Bernardi\cite{Boys:70}. 

The calculations for the graphtriyne 3D structure have been obtained by
means of density-functional theory (DFT), as implemented in the VASP code
\cite{VASP:96}, within the generalized gradient approximation (GGA) of Perdew, Burke, and Ernzerhof (PBE)\cite{pbe:96}.
The Bl\"{o}chl all-electron projector-augmented wave (PAW) method
\cite{PAW1,PAW2}, with an energy cutoff of 750 eV and a 2s$^2$ 2p$^2$ electron valence potential for carbon, has been
employed. A periodic model has been considered with an initial supercell system for
the graphtriyne isolated bilayer constituted by 48 C atoms per layer (96 C
atoms) with in-plane lattice parameter {\it a} =  12.035 \AA\,  and  {\it b} =
20.841 \AA, and with a sufficient vacuum amount to avoid spurious interaction
between bilayers along the direction normal to the {\it ab} plane.
 According to the large size of the parameters, we have sampled the Brillouin zone with a 5x4x1 Gamma centred
mesh. 
Keeping the supercell lattice parameters fixed to the initially optimized
values, and similarly keeping fixed the coordinates 
of the bottom graphtriyne monolayer,
by means of single point calculations we have progressively reduced the
interlayer distance $R$  by decreasing the coordinates along {\it c} of the top graphtriyne monolayer.

 The computed energies have been corrected by two- and three-body dispersion contributions including the Becke-Johnson
 damping scheme\cite{Grimme:11} and an Axilrod-Teller-Muto three-body term\cite{Grimme:14},  as
 implemented in the $dftd3$ program of Grimme et al.\cite{Grimme:10}. 

\section{Results and Discussion}
 
\subsection{Single layer adsorption energies}
 
In the upper part of  Fig. 1 we report the planar molecular structures that can be
considered as the smallest precursors\cite{grapheneours:2013,jpclours:2014} of
the graphene plane and of 
graphdiyine and graphtriyne pores and which are here used
 as prototypes to study their interaction with the H$_2$ molecule, shown in red 
 in the center of each structure.
In particular the potential curves obtained at the MP2C level of theory and
depicted in the lower part of Fig. 1 refer
to the H$_2$ molecule kept parallel to the molecular prototype planes and
approaching their geometrical centers. The chosen relative configurations
(sketched in the upper part of Fig. 1) are kept frozen during the calculations
and are those providing for each case both
the deepest interaction well and the lowest repulsive well at short distances. 
 In the case of graphene the minimum is located at about
 2.9 \AA\, from the plane and the related well depth is about 53 meV.
 This findings are in very good agreement with previous experimental\cite{Vidali:91} and
 theoretical\cite{dft-cc:10,dft-vdw:14} estimations. 
 Moving from graphene to graphynes the corresponding well depth
 becomes deeper and closer to the carbon plane. In particular for both graphdiyne and
 graphtriyne a well depth of about 63 meV is obtained, while
 the minimum is found at 1.8 \AA\, for the former and  right in the plane for
 the latter. 
In order to qualitatively  draw conclusions on the possibility of
 hydrogen intercalation a vertical line, corresponding to the graphite half
 interlayer distance (equals to 1.675 \AA), is also depicted.
At this distance the potential for graphene is already very repulsive (about 1000 meV) and
this clearly confirms that the existence of molecular hydrogen inside graphite
is not possible.
Nevertheless in the case of the two graphynes a quite different scenario
appears: 
for graphdiyne the minimum almost matches graphite half
interlayer distance while for graphtriyne, even if the corresponding well is located at
shorter distances, it is actually quite flat and around 1.7 \AA\, the
interaction potential is still attractive and around -50 meV.
Moreover, the penetration barrier is extremely large for graphene, it becomes quite lower for 
 graphdiyne, and it completely disappears for graphtriyne. 
Therefore from the results of Fig. 1 it can be inferred that for graphynes not only
the adsorption energy seems to be more favourable than for graphene but also that
 hydrogen out-of-plane diffusion and intercalation could be possible in multi-layer
graphynes.
     
In order to further investigate these possibilities we have studied more 
in details the features of the H$_2$--graphynes interaction by considering 
the hydrogen molecule right inside the graphynes' pores.
In fact, in Fig. 1 a low but non-negligible penetration barrier has been found
for graphdiyne and a better description of its features, which can affect the diffusion of 
 the hydrogen molecule through the pores, is desirable. 
In Fig. 2 the in-plane behavior of the H$_2$--graphynes interaction is
reported. In details, we show the interaction potential as a function of
the displacement of the H$_2$ molecule from the pore geometric center and
along two main directions (see upper  part of Fig. 2). 
First of all we notice that for both graphynes the potential does not depend dramatically
on the orientation of the H$_2$ molecule: if it is parallel to the pore plane
(solid line) then it is slightly more attractive around the geometric center than the profile for
the perpendicular configuration (dashed line); on the other hand the opposite occurs when
the potential becomes repulsive, that is for displacements in general larger than 1 \AA. 
This behaviour could be expected considering that 
the H$_2$ molecule is almost spherical from the electronic structure point of
view  since both its quadrupole moment and dipole polarizability anisotropies
are small \cite{cappelletti:08}.
More importantly, a quite large interaction anisotropy is indeed found for
graphdiyne when the H$_2$ molecule is out of the geometric center: even for
displacements as short as few angstrom tenths in both directions the potential
is more than twice
and then increase rapidly reaching very repulsive values of about 1000 meV
at around 1 \AA\, from the center. 
Therefore even if for graphdiyne the barrier for H$_2$ passage
through the pore apparently seems to be low it assumes indeed values below 50 meV just
in a very reduced area of about 0.04 \AA$^2$ around its center. These findings suggest that the
H$_2$ diffusion through the graphidiyne pore is very unlikely at standard
temperature and pressure.
On the contrary, for graphtriyne the in-pore interaction anisotropy is quite
lower: we can observe a flat well spreading in both coordinate directions and the
potential remains attractive for displacements from the center as large as about 1 \AA.
These results suggest that for graphtriyne pores there is a large area
(about 3 \AA$^2$) available for an unimpeded H$_2$ passage; it follows that
one hydrogen molecule can diffuse through the layer and be easily adsorbed in
the pore.
 We can also infer that in
principle if a further H$_2$ is taken into account both of them can not
coexist in the pore plane
 since the equilibrium distance (about 3.3 \AA) for the H$_2$
 dimer\cite{DiepJohnson:00} is larger than the diameter
 of the area describing the attractive potential.   
Therefore on the basis of the results reported in Figs.1 and 2 we can conclude
that graphtriyne sheets are better than graphdiyne ones to build up ideal carbon-based layered surfaces
suitable for H$_2$ physical adsorption; in fact, the pores of the former provide a
large area featuring an attractive interaction potential which also extends out  
 of the plane maintaining negative values lower than -30 meV
 even for distances as far as 2.5 \AA\, from the layer. 

\subsection{Multi-layer structure and adsorption energies}

In order to further investigate the possibility of hydrogen adsorption on
graphtriyne surfaces the next convenient step has been to consider a
multi-layer to assess whether intercalation is really energetically favorable or not. 
To do that we have considered a novel kind of graphite which is composed of stacked graphtriyne
sheets instead of graphene layers; its actual 3D structure represents a critical
point since it could occur that the interlayer distance is too short to allow
the hydrogen molecules to be hosted and also that the pore of one layer is
partially obstructed by benzenic rings of the adjacent layers.

The 3D arrangement of bulk graphyne and graphdiyne has been recently
investigated\cite{bulk-graphyne:2013,h2-graphd:2013,bulk-graphd:2013}  by
means of periodic dispersion-corrected DFT calculations and it has been found
that the corresponding interlayer equilibrium distance is likely
larger\cite{h2-graphd:2013,bulk-graphyne:2013} than that of graphite.
Nevertheless we believe that the methodologies employed in these studies are not accurate enough
to guarantee the reliability of the estimated 3D structure since the reported
results in the case of pristine graphite, which is used as a
benchmark test, showed that the related interlayer equilibrium distance
is not correctly reproduced. 
Therefore our choice has been instead to employ for the graphtriyne multi-layers the approach very recently introduced by Brandenburg et
al. \cite{Brandenburg:2013} which provided accurate and reference values for both exfoliation energy and interlayer
separation of graphite.

In particular we have considered at first a supercell containing
an isolated  graphtriyne bilayer (see Fig. 3) and we have assumed that its relative configuration is
similar to that of the most stable Bernal stacking found in
pristine graphite: benzene rings in different adjacent layers are displaced 
 of 1.431 \AA\, along one of the acetylenic chains joining two contiguous rings. 
 This relative orientation corresponds to a minimum also in the case of bulk graphdiyne\cite{bulk-graphd:2013}
and it has been considered frozen throughout the
DFT calculations whose results provided the cell energy as a function of the
interlayer distance, as shown in Fig. 3.
We have found that the equilibrium interlayer distance for the
graphtriyne bilayer is 3.45 \AA, which is just 0.1 \AA\, larger than that 
for pristine graphite. This result could be understood
considering that a graphtriyne plane contains a lower
density of carbon atoms with respect to graphene and consequently the size of both
attractive and repulsive contributions to the interlayer interaction is presumably reduced.


Once the structure of the graphtriyne bilayer has been determined we have used
the related equilibrium geometry to build up a graphtriyne multilayer prototype.
To do that we have considered three parallel graphtriyne pores in a
Bernal-like stacking and with adjacent layers separated of 3.45 \AA. 
The used prototype
for the graphtriyne trilayer is reported in Fig. 4 and its choice is
justified since its reduced size allows the
application of the accurate but computationally expensive MP2C level of
theory to determine the H$_2$ adsorption energy. In Fig. 4 the positions
 of the  H$_2$ molecule considered for the calculations
 are also reported and they are indicated with  A, B,
 C, B' and A' blue letters. In particular the A, A' and C sites correspond to in-pore configurations
 in which the  H$_2$ molecule is right in the pore geometric center and
 parallel to the layers. B and B' equivalent sites correspond instead to the H$_2$ intercalation:
 the molecule center of mass lie right in the direction (dashed line) joining
 the geometric centers of adjacent pores and it is placed  at 1.725 \AA\, (half the interlayer
 distance) from the closest layers, with the diatom oriented parallel to them.
The reported adsorption energies obtained at the MP2C level and shown in the lower part of Fig. 4  correspond to the sum
of three contributions related to the H$_2$ molecule interacting with each of the
 graphtriyne pore.   
It can be seen that for the in-pore A, A' and C locations the obtained interaction
energies are significantly larger than that for the single pore case (see
Figs. 1 and 2): clearly the largest interaction correspond to the C case with the 
 H$_2$ molecule inside the intermediate pore and the related adsorption energy is
 about 96 meV. 
As for the B and B' intercalation sites, the interaction energies are also very
attractive, around -90 meV, even if slightly less than that for the C position.
The obtained results for the C and B sites should be considered as reliable estimations of
typical in-pore and intercalation interaction energies for H$_2$ 
inside a periodic graphtriyne multi-layer and their large value, especially considering that only non-covalent interactions
 are involved, allow to suggest
 the following conclusions.
The first point is that no impediment is present for H$_2$ penetration across
graphtryine multi-layers since it can pass from a pore to the adjacent one
along the $z$ direction (see Fig. 4) without any barrier. 
This means that the out-of-plane diffusion through the
novel porous graphite is feasible. Nevertheless it should be clarified that in-plane
diffusion (i.e. along the directions perpendicular to $z$)
 is instead very unlikely for the inner layers since the interaction energy becomes very repulsive when H$_2$ is close 
to the carbon atoms (see Fig. 2 and Fig. S1 of the Supplementary Material).
The second point to highlight is that the quite favourable 
interaction energies imply that this novel porous
 carbon material could be used for hydrogen reversible storage.

On the basis of present results we can safely conclude that at least one hydrogen
molecule could be stored per graphtriyne pore; if we consider the periodic
cell reported in Fig. 3 we can qualitatively estimate gravimetric and volumetric
capacities of about 1.4 wt\% and 0.015 kg/L$^{-1}$, respectively. Even if these predictions are
comparable with the best hydrogen uptakes obtained so far for other carbon
nanostructures\cite{nanotubesexp:05,carbonnanostruc:15} at room temperature,
they are however somehow far from the best target proposed by the DOE\cite{DOE-note}.
Still, by taking into account that intercalation is also possible, 
 we expect that more than one hydrogen molecules could 
accommodate themselves above and below each pore. As an example, if we assume
that three H$_2$ molecules can coexist between the parallel openings we could reach  gravimetric and volumetric
capacities of about 4 wt\% and 0.046 kg/L$^{-1}$, respectively, which are in
the range or even beyond the wanted storage limits\cite{DOE-note}.   
 
\section{Conclusions}

In summary, by means of electronic structure computations,
 we have shown that graphynes are more suited than graphene for 
 the physical adsorption of H$_2$.
As a matter of fact the adsorption energy on a single layer is larger on
graphynes and the H$_2$ equilibrium distance is closer to the carbon plane.
The features of the considered  graphynes pores have been further
investigated and we have demonstrated that for graphdiyne the in-pore interaction is repulsive 
 and represents a high impediment to H$_2$ crossing; on the contrary for
 graphtriyne the in-pore interaction is attractive and a flat minimum is
 found.
 On the basis of these results a novel type of graphite, composed of
 graphtriyne stacked layers, is proposed and its structure is determined.
 By performing calculations of the interaction energy for the graphtriyne multi-layer 
 we have found that the H$_2$ molecule can freely diffuse in the direction
 perpendicular to the carbon sheets and that in-pore and intercalation sites 
 provide very favourable adsorption energies in the range of 0.1 eV.
 
Present data provide reliable benchmark results that can be used to
tune-up a new force field suitable to perform dynamical simulations
 capable to correctly assess the actual H$_2$ uptake of this novel material. 
Work in this direction is in progress.

The proposed porous graphite could be considered as a promising 
alternative to more traditional adsorbing materials such as zeolites and MOF;
in fact it shares with them a cristallyne structure with regular pores of well 
defined size but with the vantage points of being in principles lighter, chemically inert and
thermally stable. Hopefully theoretical investigations on this topic can further
stimulate the progress in the synthesis of large pore graphynes such as graphtriyne, the building
block of the porous graphite here introduced.

\section*{Acknowledgments}

The work has been funded by the Spanish grant FIS2013-48275-C2-1-P. 
Allocation of computing time by CESGA (Spain) 
is also  acknowledged.

\section*{Supplementary Material}
  Intermolecular potentials related to different configurations
  used for graphtriyne pore--H$_2$ computations are reported in an
  additional figure.
  This information is available free of charge via the Internet.



\begin{figure}[h]
\includegraphics[width=9.cm,angle=0.]{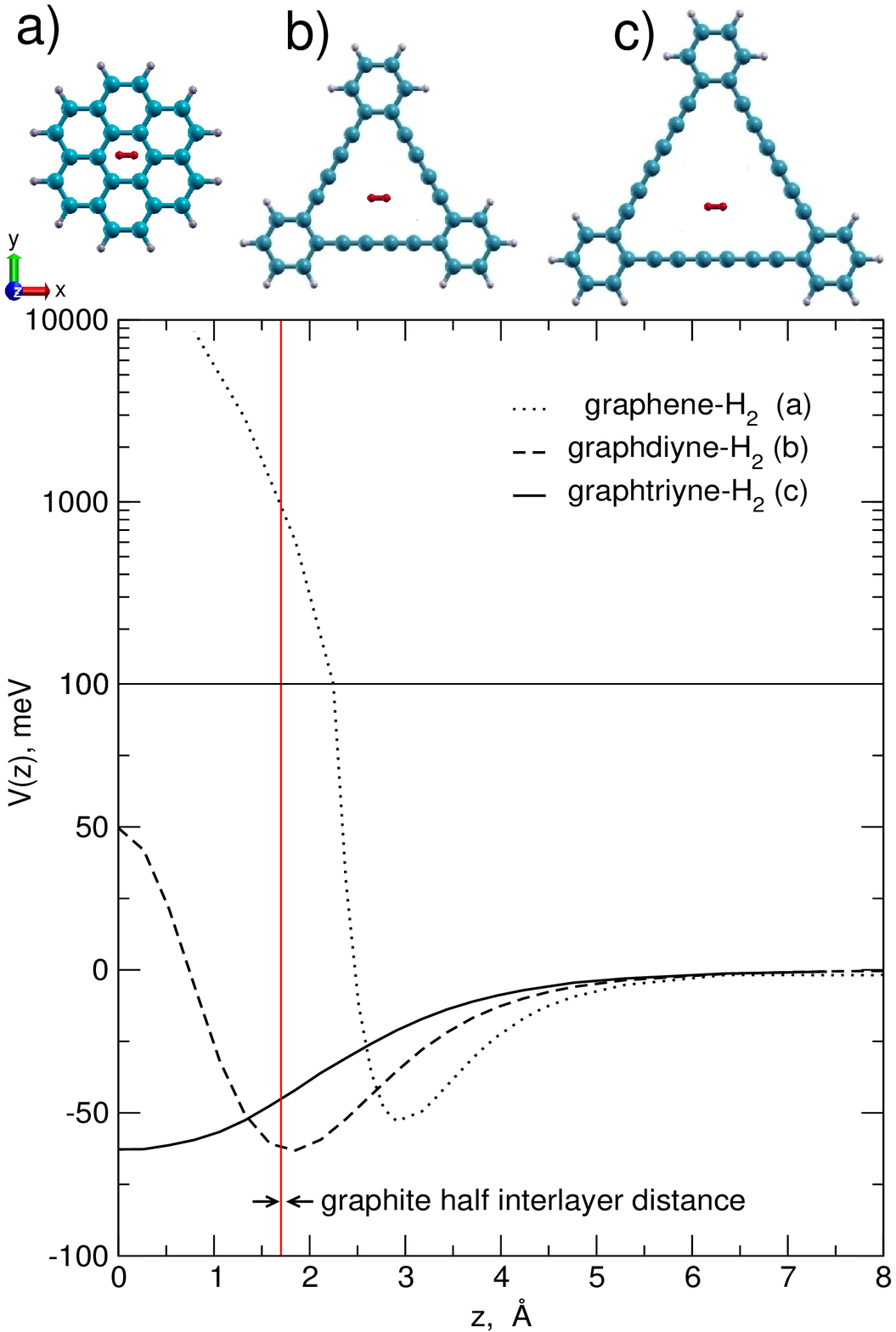}
\caption[]{ Upper part: planar molecular prototypes used 
 to study the graphene plane (a) and the pores of graphdiyne (b) and
 graphtriyne (c); they are known as coronene,
 dodecadehydrotribenzo[18]annulene and 
octadecadehydrotribenzo[24]annulene, respectively. The H$_2$ molecule is
shown in red, and its center of mass coincides in each case with the geometric center of the
coronene and annulenes; their relative distance is indicated as $z$.
 Lower part: adsorption energy profiles  obtained at the MP2C level of theory  for H$_2$
 approaching the geometric center of the planar molecular prototypes and kept
 parallel to them. The vertical red line placed at 1.675 \AA\, corresponds to the graphite half interlayer distance.}
\label{fig1}
\end{figure}

\begin{figure}[t]
\includegraphics[width=9.cm,angle=0.]{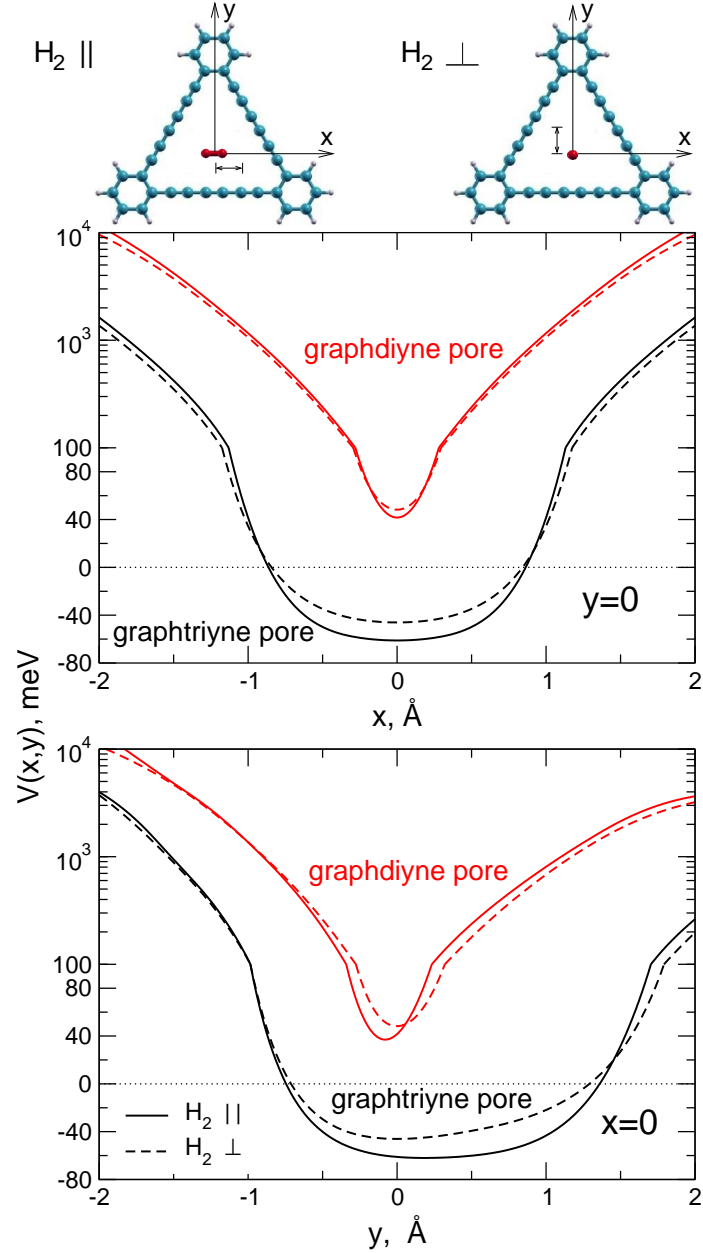}
\caption[]{In-plane behavior of the H$_2$--graphdiyne(red curves) and
  H$_2$--graphtriyne (black curves) interactions as functions of the H$_2$ displacements ($x$
  and $y$) from the geometric center of the pores. Two different H$_2$
  orientations are considered, namely parallel (solid lines) and
  perpendicular (dashed lines) to the
  annulenes' planes. The black double-headed arrows shown inside the
  triangular pores roughly describe a
  linear distance of about 2 \AA\, in the $x$ or $y$ directions.
Upper panel: $y$ is kept equals to 0 while $x$ may vary.
Lower panel: $x$ is kept equals to 0 while $y$ may vary.}
\label{fig2}
\end{figure}

\newpage

\begin{figure*}[t]
\includegraphics[width=12.cm,angle=0.]{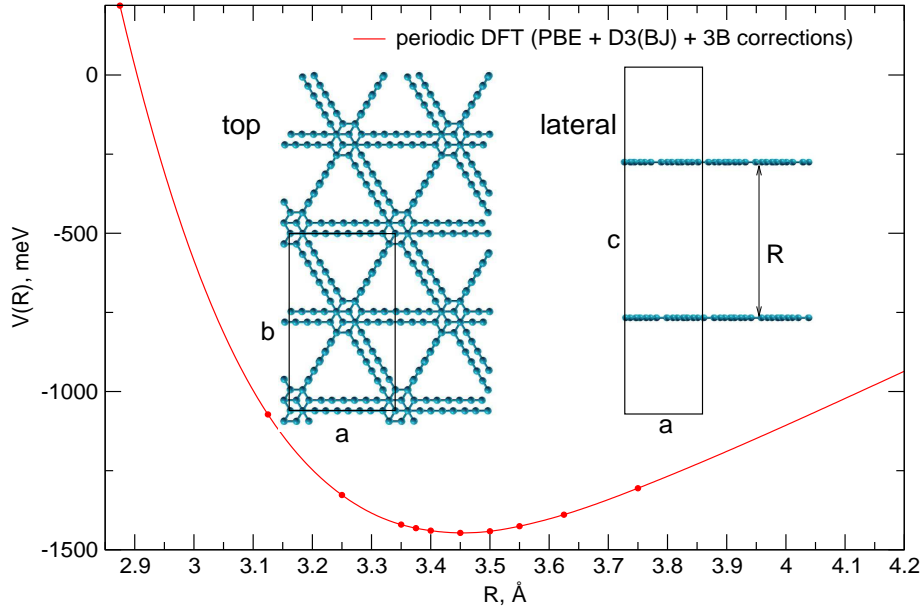}
\caption[]{ Top and lateral view of the  supercell describing the graphtriyne bilayer used 
 in the periodic DFT calculations; {\it a}, {\it b} and {\it c} lattice parameters are also
 reported and throughout the calculations they are kept frozen while the
 interlayer distance $R$ is varied. The DFT cell energies (relative to the
 asymptotic layer separation)
 are properly corrected by an ``a posteriori'' dispersion correction which take into account for
 the Becke-Johnson (BJ) damping scheme and three-body contributions (3B) and then depicted as a function of $R$. }
\label{fig3}
\end{figure*}

\begin{figure*}[t]
\includegraphics[width=11.5cm,angle=0.]{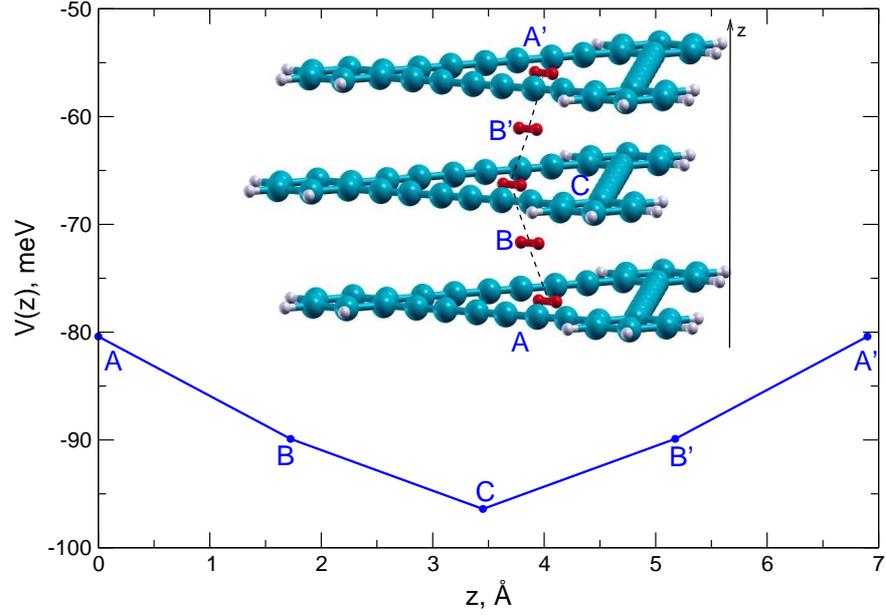}
\caption[]{ Interaction energy evolution of one H$_2$ molecule crossing a porous
  graphite composed of stacked graphtriyne layers. 
 A prototype consisting of three parallel
  graphtriyne pores in a Bernal-like 3D  arrangement is used.  The A, A', B, B' and C letters indicate
  different adsorption sites of the H$_2$ molecule: A, A' and C locations correspond to in-pore configurations
 while B and B' sites to intercalation equivalent positions. The
 five adsorption sites lie on the dashed lines which join the
 geometric centers of adjacent pores and the layers separation is fixed at 3.45 \AA, the equilibrium distance provided by
 the calculations shown in Fig. 3. }
\label{fig4}
\end{figure*}

\clearpage
\newpage







\end{document}